\documentclass[aps,prd,reprint,amsmath,amssymb,mathptmx,floatfix,footnote]{revtex4-2}
\usepackage{graphicx}
\usepackage{xcolor}
\usepackage{color}
\usepackage{bm}
\usepackage{float}
\begin{document}
\title [Quasi neutron-vortex scattering and pulsar glitches]
{Glitches due to (quasi) neutron-vortex scattering in the superfluid inner crust of a pulsar}
\author{Biswanath Layek}
\email{layek@pilani.bits-pilani.ac.in}
\author{Deepthi Godaba Venkata}
\email{godabadeepthi@gmail.com}
\author{Pradeepkumar Yadav}
\email{yadavpradeep878@gmail.com}
\affiliation{Department of Physics, Birla Institute of Technology and Science, Pilani 333031, India}
\date{today}
\begin{abstract}
We revisit the mechanism of vortex unpinning caused by the neutron-vortex scattering \cite{prad1}
in the inner crust of a pulsar. The strain energy released by the crustquake is 
assumed to be absorbed in some part of the inner crust and causes pair-breaking quasi-neutron 
excitations from the existing free neutron superfluid in the bulk of the inner crust. 
The scattering of these quasi-neutrons with the vortex core normal neutrons unpins a large 
number of vortices from the thermally affected 
regions and results in pulsar glitches. We consider the geometry of a cylindrical 
shell of the affected pinning region to study the implications of the vortex unpinning 
in the context of pulsar glitches. We find that a pulsar can release 
about $\sim 10^{11} - 10^{13}$ vortices by this mechanism. These numbers are equivalent to 
the glitch size of orders $\sim 10^{-11} - 10^{-9}$ for Vela-like pulsars with the
characteristic age $\tau \simeq 10^4$ years. We also suggest a possibility of a vortex 
avalanche triggered by the movement of the unpinned vortices. A rough estimate of the 
glitch size caused by an avalanche shows an encouraging result.
\end{abstract}
\maketitle
%%%%%%%%%%%%%%%%%%%%%%%%%%%%%%%%%%%%%%%%%%%%%%%%%%%%%%%%%%%%%%%%%%%%%%%%%
\section{Introduction}
\label{section:intro}
After the discovery of a pulsar glitch in the Vela pulsar in 1969 \cite{radh69}, a large number of 
glitches have been observed \cite{espi11} and reported to date \footnote{Updated catalogue of 
pulsar glitches : {\textit http://www.jb.man.ac.uk/pulsar/glitches/gTable.html}}. 
The size of glitches lies in the range $\sim (10^{-5} - 10^{-12})$, with an interglitch time 
of a few years. A sudden change of moment of inertia (MI) of a pulsar caused by a crustquake 
was initially proposed \cite{rude69} to be responsible for such events. However, it is now realized 
\cite{baym71} that though the crustquake model can account for small size 
crab-like glitches $\Delta \Omega/\Omega \le 10^{-8}$, the model is not compatible with Vela-like 
glitches ($\Delta \Omega/\Omega \simeq 10^{-6}$). Presently, the models associated with 
pinning and unpinning of superfluid vortices \cite{ande75,rude76} are considered 
to be the most popular models for glitches. Although the crustquake model \cite{rude69}
is not compatible with large size glitches, the crustquake event is believed to be a regular 
phenomenon of a rotating neutron star. Hence, there were suggestions to relate crustquake as a
source of other astronomical events, viz., the giant magnetic flare in magnetars 
\cite{thom95,land15}, observed gamma-ray burst \cite{gama76}, 
the emission of gravitational waves \cite{keer15,prad2} from isolated pulsars, etc. From the 
glitch perspective, 
there were attempts to unify the crustquake model with the model of superfluid 
vortices \cite{mela08, eich10}. For example, the authors of Refs.\cite{mela08, eich10} 
proposed that the crustquake might act as a trigger mechanism for vortex avalanches, which
is responsible for the release of a large number of vortices 
($\sim 10^{18}$) from the inner crust of the star and hence can produce large size glitches. Similarly 
in Ref.\cite{akba17}, it was suggested that the motion 
of vortices attached to the broken crustal plate caused by the crustquake might be responsible 
for the glitches. There was also a study \cite{epst96} of glitches through thermal creep theory, where 
the pulsar glitch was suggested to be driven by sudden energy deposition in the inner 
crust. The crustquake has been assumed to be one such resource for energy deposition. The deposited 
energy propagates as thermal waves throughout some parts of the inner crust and raises the temperature
locally. This affects the coupling between the neutron superfluid and the rigid outer crust, causing 
the star to spin up.

The above discussion suggests that besides explaining small size glitches or attempting to understand a 
few astronomical events, the crustquake is also believed to have a role in the superfluid vortex model.
In view of this, some of us \cite{prad1} earlier proposed a mechanism of vortex unpinning initiated 
by a crustquake. The purpose of the unification of the crustquake with the superfluid vortex model is 
to explain large size glitches without changing the interglitch time. Note that in 
the crustquake model, a large glitch requires a longer interglitch time, contrary to the observation. 
Thus, the model alone is not sufficient for explaining large size glitches. On the other hand, 
though the superfluid model is compatible with the larger glitches, it requires a dynamical 
mechanism to unpin the vortices from the inner crust of the star. The occurrence of a 
crustquake followed by vortex unpinning may resolve the above important issue associated with 
the size and frequency of glitches from an individual pulsar. 
With this aim, we will further explore the unpinning mechanism as proposed 
in Ref.\cite{prad1} by supplementing a few more qualitative arguments in favor of 
the unpinning process. In Ref.\cite{prad1}, the study was initiated by taking a simple cubical 
geometry of the affected pinning site to test mainly the new vortex unpinning mechanism. Here we 
will extend our analysis by considering a cylindrical geometry of the affected pinning region. 
Such geometry is taken, partly encouraged by the studies of thermal glitches through creep 
theory in Ref.\cite{epst96}, where authors assumed an energy deposition in a cylindrical 
shell in the inner crust region. However, the shell was taken in an arbitrary location in that 
study. As the location of the pinning site (i.e., the place where the thermal energy is presumably 
absorbed) was unknown, we will study the effect by taking the shell at different locations in 
the inner crust. In this case, depending on the affected region's location, vortex unpinning 
through avalanche is further possible. In contrast, the possibility of vortex avalanches was 
ineffective in the earlier study in Ref.\cite{prad1}. As vortex avalanches may be responsible 
for large-size glitches, it is worth exploring the such possibility in the new proposed picture 
of vortex unpinning. Here, we should mention that the energy absorption by neutron star matter 
is an important issue, and the absorption should depend on the properties of the matter in the 
inner crust, as noted by the authors in Ref.\cite{fabi76}. However, the author in that work 
assumed a simplistic polytropic equation of the state of the matter. So, working in this direction 
is essential to settle the question of energy absorption by taking a more realistic equation of 
state. However, such a study needs particular attention, and we would like to explore it in our 
future work.

We assume that the strain energy released by a crustquake is absorbed in the inner crust 
of the pulsars. This breaks a significant fraction of the neutron Cooper pairs from the existing free 
neutron superfluid in the inner crust. The creation of {\it free electrons} and 
holes by breaking up the  cooper pairs have been often mentioned in the literature \cite{nobel, pair1}
in the context of electrical superconductivity. The free electrons resulting
from the pair breaking are referred to as quasi-particles. The breaking (and formation) of Cooper pairs 
from the neutron superfluid has also been extensively discussed in the context 
of the cooling mechanism of neutron stars through neutrino emission \cite{Kolo08, pair1, pair2}. 
The existence of the superfluid gap sets the minimum energy requirement for breaking up Cooper pairs.
For the case of neutron superfluid, the quasi-particles are the neutrons, 
and the corresponding gap parameter for the bulk neutron superfluid matter 
will be denoted by $\Delta_f$. 

Here we should mention that the above picture of quasi-particle excitations 
created through the breaking of Cooper-pairs is different from the excitations 
(i.e., phonon, roton, kelvon, etc., depending on the nature of the superfluidity) as suggested through  
Landau's phenomenological two-component superfluid model \cite{landau}. 
Here we will take the
picture of quasi-neutron excitations produced from pair breaking of neutron superfluid. These quasi-neutrons 
scatter with the vortex core {\it normal} neutrons and share their energy with the pinned vortices.
(Here, the quasi-neutrons should be treated like normal neutrons. See the comments made in \cite{nobel}). 
This causes unpinning of a large number of pinned vortices ($\sim 10^{13}$) from the affected region 
in the inner crust and results in the glitch event. In the previous work \cite{prad1}, the volume of 
the pinned region, the number of vortices, and eventually the glitch size was estimated for a fixed value 
of Fermi momentum ($k_f$). In this work, all the above quantities have been evaluated by varying the Fermi 
momentum of the bulk superfluid neutrons, which can be achieved by taking the shell at a different location.
We will see an important implication if the affected shell is taken into the deeper region of the inner crust. 
In that case, the vortex unpinning from that region may trigger a vortex avalanche and results in Vela-like 
large-size glitches. We will explore this feasibility by taking a picture of proximity knock-on as described 
in the literature \cite{wars12, wars13}.

The paper is organized in the following manner. After the initial set-up of the formulations in 
section \ref{section:basic}, we will describe the geometry of the affected pinned vortex region in 
section \ref{section:geometry}. There, we will develop the mathematical tools for estimating 
various quantities, namely, thickness ($\delta_s$) of the shell, number of unpinned vortices
($N_v$), and size of pulsar glitches $\Delta \Omega/\Omega$ etc. The mechanism of 
vortex unpinning by neutron-vortex scattering will be presented in section 
\ref{section:mechanism}. We will present our results in section \ref{section:result}. The feasibility 
of the vortex avalanche through the process of proximity knock-on will be discussed in section 
\ref{section:avalanche}. We will comment and conclude our work in section \ref{section:conclusion}.

%%%%%%%%%%%%%%%%%%%%%%%

\section{Pulsar Glitches Through Vortex Unpinning : The basic Formulation}
\label{section:basic}
First, we will mention the neutron star model, based on which 
we will implement the vortex unpinning mechanism \cite{prad1} for studying pulsar glitches. 
Since the possibility of the existence of neutron stars was hypothesized in 1934 
\cite{Baad1,baad2}, there have been attempts to understand various properties 
of the neutron star. For a given mass of a neutron star, the internal structure (such as 
the inner crust thickness, its moment of inertia, etc.)  and the bulk properties (such as 
the radius, total moment of inertia, etc.) depend on the equation of state (EOS) of the 
nuclear matter \cite{nege73,mark96,link99, stei13}. However, 
extracting the precise values of neutron star parameters is still challenging, mainly because 
of non-adequate knowledge of EOS for the highly dense matter prevailing inside the star. For our 
purpose, we will take a broad 
picture of the neutron star's internal structure that has emerged after several studies 
and is taken in the studies of various pulsar phenomena. A neutron star of mass about 
$1.4 M_\odot$ and radius of order (10-12) km consists of the inner crust of thickness 
(1-2) km sandwiched between (0.3 - 0.5) km thick rigid outer crust and the core of radius 
$\sim 9$ km \cite{mark96,link99, stei13}. The baryon density in the inner crust for 
such a stellar mass neutron star lies in the range $(10^{11} - 10^{14})~\text{g-cm}^{-3}$. 
The matter at such high baryon density exists in the form of neutron superfluid \cite{migd59,ande75}, 
even though the internal temperature can be in the sub-MeV scale \cite{epst96}. The denser inner core 
consists of quantum liquid forming neutron and proton superfluid. There is a conjecture for 
the existence of an even more exotic form of matter inside the core of a more massive and compact 
neutron star \cite{latt04}. For such a case, the values of various parameters are expected to deviate 
from those mentioned above. As our focus will be mainly on Vela-like pulsars, we will assume 
the neutron stars of mass $\sim 1.4 M_\odot$ with radius $\sim$ 11 km \cite{mark96} and study the pulsar 
glitches implementing the vortex unpinning picture of \cite{prad1} in the standard superfluid vortex 
model \cite{ande75}. So, we will 
also fix the other neutron star parameters, most importantly, the thickness of the inner crust,
based on the above neutron star model only. For more 
precise calculations and consistency checks, one should extract the information about the neutron star 
parameters by choosing the EOS correctly. We will discuss the issue of parameter-dependent sensitivity 
of our results in an appropriate place (section \ref{section:result}).

For a rotating neutron star, the rigid part 
of the star's crust (i.e., outer crust) and the inner core are assumed to corotate with a moment of 
inertia $I_c$. The neutron superfluid component of a moment of inertia $I_f$ in the inner crust acts 
as an angular momentum reservoir in the form of pinned vortices. 
The time $t = 0$ is set when the superfluid vortices are pinned in the inner crust region. After a 
crustquake, a fraction of these vortices get unpinned \cite{prad1} at $t = t_p$ and results 
in pulsar glitches. We will see that the 
unpinning occurs from a localized region of the inner crust, and we call it {\it local unpinning} to distinguish 
the unpinning occurring through the avalanche process (section \ref{section:avalanche}).
 The geometry of the affected region and the unpinning mechanism will be discussed in section 
\ref{section:geometry} and \ref{section:mechanism}, respectively. The time $t_p$ is the interglitch time and 
assumed (such assumption will be justified later) to be of the same order as the frequency of the successive 
crustquakes. We denote $\Omega_p$ as the angular velocity of the superfluid component in the inner crust, 
which remains fixed during $t = 0$ to $t = t_p$. $\Omega_c (t)$ is the angular velocity of the co-rotating 
crust-core system with $\Omega_c (0) = \Omega_p$. The development of differential angular velocity 
$\delta \Omega = \Omega_p -\Omega_c (t)$ between the inner crust superfluid and the rest of the star follows 
the time evolution of the star. The differential angular velocity at $t = t_p$ can be expressed as 
\begin{equation} \label{eq:domega1}
\frac{\Omega_p - \Omega_c (t_p)}{\Omega_c (t_p)} \equiv \Big(\frac {\delta \Omega} {\Omega}\Big)_{t_p} 
= \frac {t_p}{2\tau}.
\end{equation}
Where $\tau = - (\Omega/2 \dot \Omega)$ is the characteristic age of the pulsar and we assume, $t_p << \tau $. For the 
ease of notation, from now onward, we denote $(\delta \Omega/\Omega)_{t_p}$ as $\delta \Omega/\Omega$. 
Applying the standard superfluid vortex model \cite{ande75,rude76} (see also the review \cite{hask} 
for various models of glitches), 
the glitch size can be written as 
\begin{equation} \label{eq:Domega1}
\frac{\Delta \Omega}{\Omega} = \Big(\frac{I_f}{I_c}\Big) \Big(\frac{\delta \Omega}{\Omega}\Big) 
\Big(\frac{N_v}{N_{vt}}\Big) = \Big(\frac{I_f}{I_c}\Big) \Big(\frac{t_p}{2\tau}\Big) 
\Big(\frac{N_v}{N_{vt}}\Big) .
\end{equation}
Here $I_f/I_c$ is the MI ratio of the bulk neutron superfluid in the inner crust to the rest of the star. 
The quantities $N_{v}$ and $N_{vt}$ are the number of pinned vortices in the 
affected region and the total number of pinned vortices in the equatorial plane in the inner 
crust (Eq. (\ref{eq:vortex}) and Eq. (\ref{eq:nvt})), respectively. The ratio 
$N_v/N_{vt}$ takes care of the fact that only a fraction of the pinned vortices is affected by 
the local unpinning.
%%%%%%%%%%%%%%%%%%%%%%%
\section {Mathematical tools for Estimating various quantities related to glitch }
\label{section:geometry}
As proposed in Ref.\cite{prad1}, the thermally excited neutrons are responsible for 
unpinning the superfluid vortices. In turn, the number of unpinned vortices depends on the release of 
energy in a crustquake event. Here we will briefly describe the crustquake model
\cite{rude69}, mainly focusing on the model's essential features relevant to our 
study (for more details, see the Ref.\cite{baym71}). The oldest 
theoretical model for pulsar glitches, namely, the crustquake model, assumes the existence of a solid 
deformed crust of a pulsar. The oblateness parameter $\epsilon$ can characterise the deformation 
as $\epsilon = \frac{I_{zz} - I_{xx}}{I_{0}}$. Where $I_{zz}$, $I_{xx}$ and $I_0$ are 
the moment of inertia about the z-axis (rotation axis), x-axis, and the spherical star, respectively 
\cite{baym71}. At an early stage of formation, with a very high rotational frequency, the crust 
solidified with an initial larger value of oblateness. As the star slows down due to electromagnetic 
radiation loss, the oblateness $\epsilon (t)$ keeps decreasing, causing crustal strain development 
in the star's outer crust. Once the critical strain is reached, the crustquake occurs to achieve 
a new equilibrium. This results in sudden change (decrease) of oblateness $\Delta \epsilon$. 
Henceforth, we will take $\Delta \epsilon$ to be positive. The decrease in oblateness
causes the star's moment of inertia (MI) to decrease, increasing its rotational frequency 
(following angular momentum conservation). The glitch size ($\frac {\Delta \Omega}{\Omega}$) of a 
pulsar is directly related to $\Delta \epsilon$ through $ \frac {\Delta \Omega}{\Omega}= - 
\frac {\Delta I}{I_0} = \Delta \epsilon$. Immediately after the proposal, it was realized \cite{baym71} 
that the interglitch time or the waiting time of two successive glitches is also determined by the 
change of oblateness and is proportional to $\Delta\epsilon$. As noted in Ref.\cite{baym71}, $10^{-8}$ 
size Crab-like glitches requires an average one-year waiting time \cite{lyne15}. Thus the glitch size 
being proportional to $\Delta\epsilon$, a larger glitch needs a longer waiting time, contrary to 
the observations. For example, for two successive glitch events, Vela pulsar (of glitch size 
$\simeq 10^{-6}$) requires about 100 years. This is the most critical problem the crustquake model 
has encountered since its inception. As mentioned before, unifying the crustquake with the superfluid 
vortex model may explain large-size glitches with compatible interglitch time. Now, for determining 
the energy release in a single crustquake event,  we will take $\Delta\epsilon = 10^{-8}$. This 
value is consistent with a typical one-year observed waiting time between successive glitch events. 
As the crustquake occurs, it releases strain energy of magnitude $\Delta E = B \Delta \epsilon$. 
Here the constant $B$ ($\sim 10^{48}$ erg) is related to the modulus of rigidity of the crust. 
As mentioned earlier in section \ref{section:basic}, the released energy by the crustquake is assumed 
to be thermally absorbed in a local region in 
the inner crust of the star. This results in breaking up neutron-neutron Cooper pairs from the bulk neutron superfluid
and creates free quasi-neutrons (excited neutron) in the inner crust. The sharing of energy through the scattering of 
these excited neutrons with the vortex core normal neutrons causes the unpinning of vortices 
(section \ref{section:mechanism}) and results the glitch event. For the estimate of the number of unpinned 
vortices $N_v$, the affected pinning region is taken to be of a cylindrical shell of height $h_s$ and thickness 
$\delta_s$. Such geometry was also taken in Ref.\cite {epst96} in their study of pulsar
glitches through thermal creep theory. However, barring the geometry, our formulation 
of generation of pulsar glitches is completely different compared to the approach taken in Ref.
\cite{epst96}. The reason for choosing the affected region around the equatorial plane is motivated by 
the crustquake picture of \cite{baym71}. Also, in the studies of the crustal strain \cite{fran00}, the 
authors have found the strain to be maximum at the equatorial plane, which makes it the most likely 
place for the quake site. 

As we will see, the thickness ($\delta_s$) of the cylindrical shell that determines 
the number of pinned vortices therein crucially depends on the Fermi momentum $k_f$ of the bulk superfluid
neutrons, which in turn depends on the baryon density of the specific region. We will follow the work of 
Pastore et al. \cite{past11}, where the authors have studied the properties of the neutron superfluid 
in the inner crust of the star (see also the seminal work of Negele and Vautherin \cite{nege73}). 
In that work, the Fermi momentum was computed for the baryon density region ranging from $\rho \simeq 
10^{12}~\text{gm~cm}^{-3}$ to $\rho \simeq 10^{14}~\text{gm~cm}^{-3}$. As mentioned by the authors,
the spherical Wigner-Seitz approximation, which was used to calculate the various quantities,
can reproduce well ground-state properties of the outermost regions of the inner crust. 
However, as the methodology breaks down beyond $\rho \simeq 10^{14}~\text{gm~cm}^{-3}$ ; we will 
restrict our study up to the baryon density $\rho = 10^{14}~\text{gm~cm}^{-3}$.  The corresponding 
Fermi momentum is given as $k_f = 1.2 ~\text{fm}^{-1}$ 
(Table 1 of Ref.\cite{past11}).
As the Fermi momentum depends on the 
local mass density, or equivalently on the depth of the crust, the value of $\delta_s$ should also depend on 
the location of the shell. As the precise location of the thermally affected region is unknown to us, we will 
vary the distance of the shell $R_s$ (as measured from the center of the star) from $10.3$ km to $9.9$ km for 
the estimating of various quantities associated with the glitches. The above values of $R_s$ are in accordance 
with the Fermi momentum in the range $(0.2 - 1.2)~\text{fm}^{-1}$ and local mass density in the range 
$(10^{12} - 10^{14})~\text{gm~cm}^{-3}$ (Fig.1 of \cite{epst96} and Table 1 of \cite {past11}). 
Here we should mention that the authors \cite {past11} have calculated a finite discrete set of
Fermi momentum at various values of the mass density in the inner crust of the star. However, we
have assumed a continuous distribution of $k_f$ within a particular range as mentioned above. 
We hope that the interpolation of $k_f$ in between two successive data will not
make any significant changes to our results. Note that it is due to the uncertainty on the 
location of the shell, the various quantities are calculated at various Fermi 
momentum. However, it serves a few useful purpose of testing the sensitivity of 
the results due to the uncertainty in the affected region. As mentioned earlier, it also 
allows exploring the feasibility of avalanches through the picture of proximity knock-on 
(section \ref{section:avalanche}) triggered by the unpinned vortices.

We will now  calculate the thickness of the shell $\delta_s$, the number of the unpinned vortices $N_v$, 
and the glitch size $\Delta \Omega/\Omega$ at various values of the Fermi momentum.
The volume $V_s$ of the affected cylindrical region can be estimated by energy balance \cite{prad1,wang21} 
\begin{equation} \label{eq:energy}
B \Delta \epsilon = N_e \Delta_f = \frac{\Delta_f^2}{E_f} n_f V_s , 
\end{equation}
i.e.,  
\begin{equation} \label{eq:vp1}
V_s = \frac{B \Delta \epsilon ~E_f}{n_f \Delta_f^2}.
\end{equation}
%%%%%%%%  FIGURE -1 %%%%%%%%%
\begin{figure}
\centering
%\vspace{-1.5cm}
\includegraphics[width=1.1\linewidth]{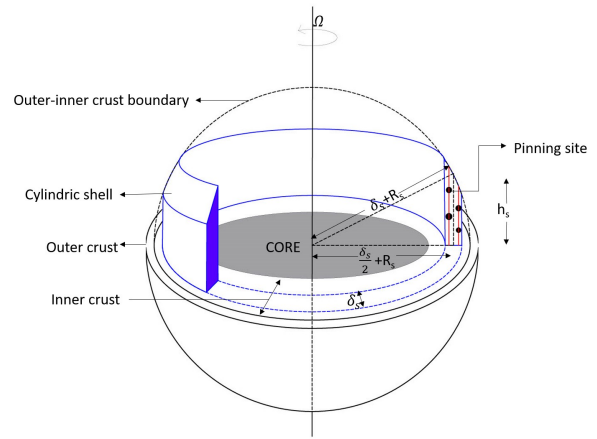}
%\vspace{-0.5cm}
\caption{The cylindrical shell of thickness $\delta_s$ and height $h_s$ represents 
the affected pinning site (blue colored region) in the inner crust. The 
vortex lines (red color) terminating on the outer-inner crust boundary defines 
the (average) height of the cylinder.}
\label{fig:fig1}
\end{figure}
Where $n_f$, $\Delta_f$ and $E_f$ denote the number density of the bulk superfluid neutrons, the energy gap 
parameter, and the Fermi energy, respectively. $N_e$ is the number of excited neutrons produced from the
Cooper-pair break up. For a cylindrical geometry, 
the volume $V_s$ within which the energy is deposited can be written as $V_s = 2\pi R_s  h_s \delta_s$. 
The number of pinned vortices 
in the enclosed area $A_s~(= 2\pi R_s \delta_s)$ in the equatorial plane can be estimated from the vortex areal
density $n_v = 4m_n \Omega/h \simeq 10^7 \text{m}^{-2}  (\Omega /\text{s}^{-1}) $. The height of the cylinder 
is determined by defining the affected pinning region such that all the vortex lines enclosed in an equatorial 
area $A_s$ will contribute to the vortex unpinning. For this, the cylinder is cut at an
approximate height $h_s$, where the vortex lines terminate on the boundary of the inner crust with the outer crust.
Here we should mention that because of the sphericity of the inner crust,  
the affected region is not expected to be in the form of a perfect right circular cylinder. However, 
considering the number of affected vortices to be enormous, we assume the geometry to be a right circular 
cylinder for simplicity. Thus the height of the cylinder 
is given by (see Fig. \ref{fig:fig1} for illustration),
\begin{equation} \label{eq:height}   
h_s =  \Big [{(R_s + \delta_s)^2 -(R_s + \delta_s/2)^2}\Big]^{1/2} \simeq ({R_s \delta_s})^{1/2} . 
\end{equation}
Here we have taken the approximation $\delta_s << R_s$. Using the above equation, the volume $V_s$ can now
be expressed in terms of the thickness of the shell as  
\begin{equation} \label{eq:vp2}
V_s = 2\pi (R_s \delta_s)^{3/2} = \frac{B\Delta \epsilon E_f}{n_f\Delta_f^2} .
\end{equation} 
Using the above equation, we can now determine the shell thickness $\delta_s$ 
by choosing the Fermi momentum $k_f$ in the range $0.2~\text{fm}^{-1} - 1.2~\text{fm}^{-1}$. Note that 
neutron density $n_f$, Fermi energy $E_f$, and the gap parameter $\Delta_f$ all are 
functions of $k_f$. In the above, we assumed the Fermi momentum to be uniform within a shell. 
However, the vortex lines emanating from the equatorial plane pass through various baryon 
density regions; hence, $k_f$ should vary along the height of the cylinder.
Including this factor in the calculation requires a detailed density profile of the inner crust region,
which depends on EOS for a given neutron star mass. We expect some uncertainty
in our estimate of $\delta_s$ and the other associated quantities by assuming a fixed $k_f$ 
(and $n_f$, $\Delta_f$, $E_f$, etc.). Within this limitation, the number of pinned vortices 
enclosed in $A_s$ can be estimated as
\begin{equation}\label{eq:vortex}
N_v = A_s n_v = \frac{V_s n_v}{h_s} = \frac{B\Delta \epsilon E_f n_v}{ n_f\Delta_f^2 ({R_s \delta_s})^{1/2}}.
\end{equation}
We have used Eq. (\ref{eq:vp2}) for the last equality in the above equation. 
The vortex lines can not terminate in the inner crust superfluid region for 
topological reasons. These can form either a close loop or should end on the boundary of the inner 
crust. Thus, the number of vortex lines within the affected region is the same as that of vortices 
enclosed in $A_s$ (ignoring the loop).
Once we fix the thickness of the cylindrical shell from Eq. (\ref{eq:vp2}) (by choosing the appropriate values 
of the parameters), we can estimate the number of unpinned vortices using Eq. (\ref{eq:vortex}). 
We will use Eq. (\ref{eq:domega1}) and Eq. (\ref{eq:Domega1}) for the estimate of size of the glitches. 
The total number of vortices $N_{vt}$ in the whole inner crust is obtained as 
\begin{equation} \label{eq:nvt}
N_{vt} \simeq (2 \pi R \Delta R) n_v .
\end{equation}
Where, $R~(\simeq 10$ km) is the average distance of the inner crust from the center of the star and 
$\Delta R~(\simeq 1$ km) 
is the thickness of the inner crust. 
In the above, we assumed a uniform 
vortex density $n_v$ throughout the crust while estimating the number of vortices. As the vortex density, 
$n_v$ is proportional to the rotational frequency $\Omega$; in principle, one can relax such an assumption. 
One can take instead a local vortex density $n_v(r)$, which depends on the distance ($r$) from the rotation 
axis. This is equivalent to assuming the radial distance-dependent rotational frequency of the various 
superfluid region. In fact, the thermal creep theory \cite{alpa84} considers such radial distance dependence 
vortex density $n_v(r) = \frac{2\Omega(r)} {\kappa} + r \frac{\partial \Omega(r)}{\partial r}$ (see Refs. 
\cite{alpa84,epst96} for the detailed formulation of creep theory). Here $\kappa = h/2m_n$ is the quantum 
vorticity with $m_n$ being the mass of a neutron. As in creep theory, the dynamics and the steady state 
behavior of the superfluid-crust couple system are realized through the motion of vortices. In such a 
scenario, if the coupling is assumed to happen locally, taking the radial dependence of angular frequency 
and hence the vortex density is essential. However, for explaining large size glitches ($\sim10^{-6}$) 
within the standard superfluid vortex model \cite{rude76}, a very large fraction of the pinned vortices 
must be released simultaneously. In that case, assuming a common angular frequency throughout a significant 
region of the superfluid components in the inner crust is natural.
The glitch size is now obtained from Eq. (\ref{eq:Domega1}) as
\begin{equation} \label{eq:glitch}
\frac{\Delta \Omega}{\Omega} = \Big(\frac{I_f}{I_c}\Big) \Big(\frac{t_p}{2\tau}\Big) 
\Big(\frac{R_s}{R}\Big)\Big(\frac{\delta_s}{\Delta R}\Big).
\end{equation}
We will use the above equation to present the results in section \ref{section:result}.
Now, we will revisit the vortex unpinning mechanism, supplementing a few more physical arguments 
in favor of our earlier proposal \cite{prad1}.
%%%%%%%%%%%%%%%%%%%%%%%
\section {Vortex Unpinning Through Neutron-Vortex Scattering }
\label{section:mechanism}
The strain energy released in a crustquake event is taken as $\Delta E = B \Delta \epsilon \simeq 10^{40}$ erg. 
We assume this energy is partially absorbed in the cylindrical shell and thermally excites the neutrons from 
the bulk neutron superfluid by breaking some fraction of Cooper pairs. The excited quasi-neutrons, in turn, 
unpin the vortices from the shell through neutron-vortex scattering. In the volume $V_s$, the absorbed energy 
can excite about $N_e$ ($\sim \frac{\Delta_f}{E_f} n_f V_s$) number of neutrons from the  bulk neutron superfluid. 
Each of these excited neutrons, on average, has an energy $E_f$. We will see that this energy is more than the 
pinning energy $E_p$ (per site) in the cylindrical shell, irrespective of the shell's location. We also note that, if 
$N_p$ ($\sim \frac{V_s}{d_v^3}$) denotes the number of pinned vortices in the shell, then the number of 
excited neutrons per pinned vortex (i.e., $N_e/N_P$) is of order $\sim (d_v k_f)^3 \sim 10^{30}$. 
Here $d_v$ ($\sim 10^{10}$ fm) is the intervortex distance and the Fermi momentum $k_f$ is of order  
$\text {fm}^{-1}$.
Thus, each pinned vortex, on average, is surrounded by approximately $10^{30}$ excited neutrons. So these neutrons
are expected to scatter with the vortex core normal neutrons to unpin the vortex. 
In the picture of unpinning through scattering, the pinning energy $E_p$ should be treated as the binding energy of
the vortex-nucleus system arising due to the vortex-nucleus interaction. The sharing of energy by the excited neutrons 
with the vortex core neutrons increases the latter's energy. Here the energy of the excited neutron acts as activation 
energy helping to overcome the pinning barrier. The neutron-vortex scattering can be represented as \cite{prad1}\\

\noindent {\it excited neutron ($\sim E_f$) + pinned vortex ($ - E_P$) $\rightarrow$ de-excited neutron ($E_f - E_p$) + 
free vortex.} \\

In above, the energy of various objects are denoted in bracket. The negative sign in front of $E_p (>0) $ 
conventionally signifies the binding energy of a pinned vortex.

\begin{figure}
\centering
\vspace{0.5cm}
\includegraphics[width=1.0\linewidth]{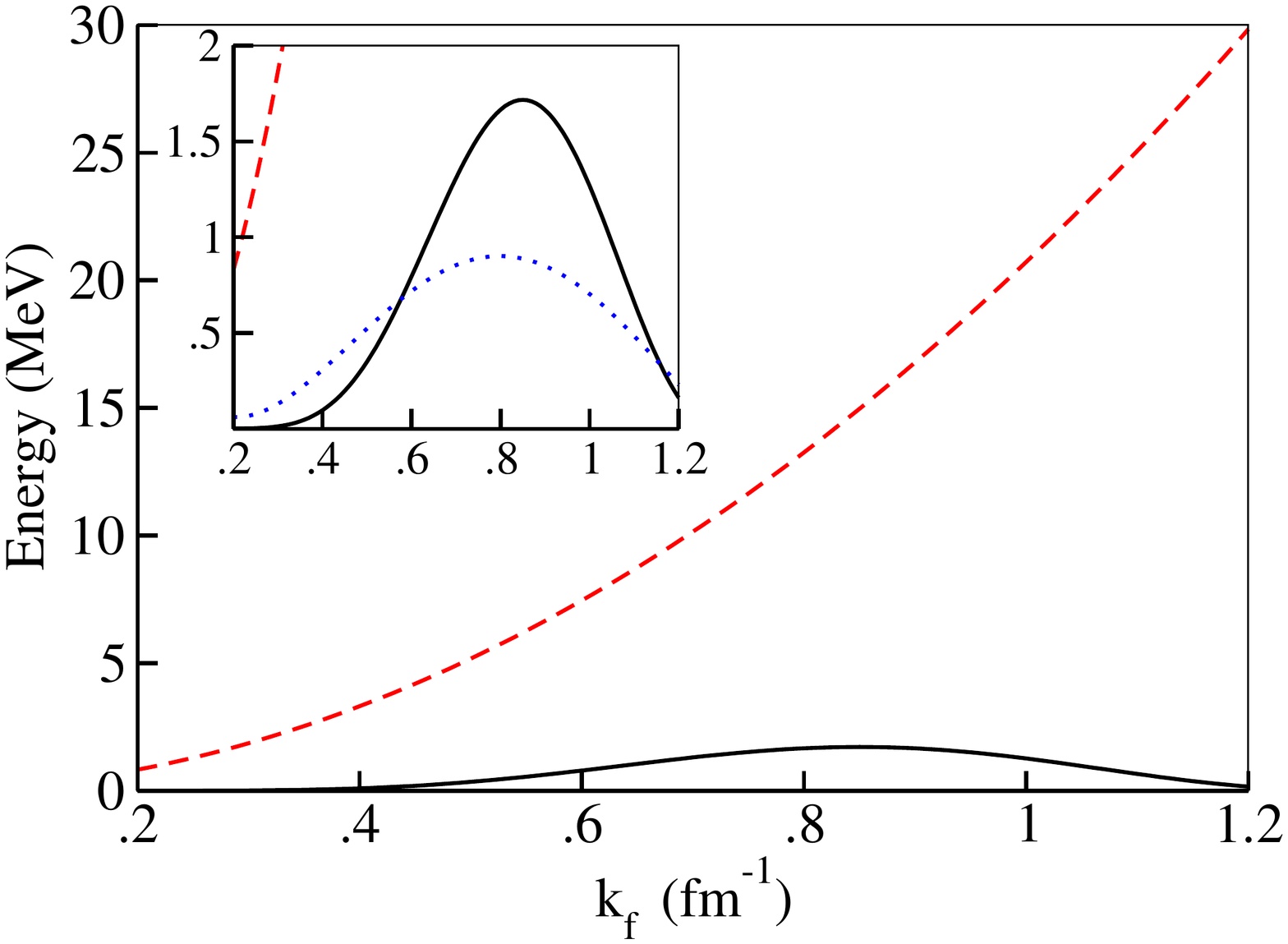}
\vspace{-1.5cm}
\caption{Excitation energy $E_f$ (red dashed) and pinning energy per site $E_p$ (black solid) as
a function of Fermi momentum. Inset shows the same plot with smaller energy range for demonstration
of $E_p$ more clearly. Blue dotted line shows the gap parameter $\Delta_f (k_f)$ (The expression 
of $\Delta_f (k_f)$ is taken from \cite{sinh15}).}
\label{fig:fig2}
\end{figure}

We now compare the average energy of the excited neutron, i.e., Fermi energy with the pinning energy. The pinning 
energy per site depends on the local mass density, or equivalently, on the Fermi momentum of the neutron
superfluid and it is given by \cite{alpa84,epst91}
\begin{equation} \label{eq:ep}
E_p = \frac{3}{8} \gamma \frac{\Delta_f^2}{E_f} n_f V.
\end{equation}
Where $V  = \frac{4}{3} \pi \xi^3$ is the overlap volume between the vortex and the nucleus. The size of a vortex 
core $\xi$ ($\simeq 10 ~\text {fm}$) is of the same order as the nuclear radius. The numerical value of $\gamma$ 
is of order unity \cite{alpa84, alpa89}. The value of $E_p$, which can be determined from Eq. (\ref{eq:ep}), 
should now be compared with the average energy (i.e., the Fermi energy) of each excited neutron. 
As per the results (see Fig. \ref{fig:fig2}), the energy required to overcome the pinning barrier is satisfied 
comfortably throughout the region of interest. 

Macroscopically, the possibility of vortex unpinning through neutron-neutron scattering can also be 
realized from the following arguments. While proposing the superfluid vortex model for glitches, 
it was \cite{ande75} mentioned that the magnitude of the frictional force arises due to the scattering 
of electrons with the vortex core neutrons (e-n scattering) is too small compared to the pinning 
force to unpin the vortices. However, as we see below,  in comparison, the frictional force that 
arises due to neutron-neutron scattering can be pretty large. To understand this,
let us write the expression of the frictional force (per unit volume) caused by n-n scattering as
(following Ref.\cite{ande75}) 
\begin{equation} \label{eq:fric1}
F_{nn} \simeq \frac{ \rho_n R~\delta \Omega}{\tau_{nn}} 
\simeq (\frac{\Delta_f}{E_f} \rho_f) ( \frac{R~\delta \Omega}{\tau_{nn}} )
\end{equation}
Where $\rho_n (\simeq \frac{\Delta_f}{E_f} \rho_f)$ and $\rho_f$ are the mass density of excited neutrons
and superfluid neutrons, respectively. The time scale $\tau_{nn}$ is set by the scattering events of the 
excited neutrons with the normal vortex neutron (neutron-neutron scattering). 
Note that for the frictional force due to electron-neutron scattering, $\rho_n$ and $\tau_{nn}$ 
will be replaced by 
electron mass density $\rho_e$ and the relevant time scale $\tau_{en}$, respectively. The numerical value 
of $\tau_{nn}$ is expected \cite{prad1} to be quite small of order $10^{-5}~ \text{s}$ compared to e-n scattering time 
scale $\tau_{en}$. Note, $\tau_{en}$ is the typical spin-up decay time of order few months \cite{baym69, ande75}. 
The small value of $\tau_{nn}$ arises solely due to larger strength of neutron-neutron magnetic moment 
interaction relative to the strength of e-n interaction \cite{baym69,prad1}. Now for 
comparison with the frictional force caused by e-n scattering, we take the free electron mass density
\cite{ande75} in the inner crust $\rho_e \simeq 10^7~\text{g-cm}^{-3}$ and $\tau_{en} \simeq 10^{7}~\text{s}$.
Similarly, for n-n scattering the mass density of excited neutron can be approximately taken as 
$\rho_{n} \simeq 10^{12}~\text{g-cm}^{-3}$. For an order of magnitude estimate, Fermi momentum 
$k_f$, for example, is taken as $0.8~\text{fm}^{-1}$ and the values of other parameters are 
fixed accordingly. Thus there is about $10^{17}$ order of magnitude enhancement of the frictional force due to
n-n scattering in comparison to e-n scattering. The frictional force for n-n scattering turns 
out to be quite large because of larger values of $(1/\tau_{nn}) \simeq 10^{12} (1/\tau_{en})$ 
and $\rho_n \simeq 10^5 \rho_e $. This makes the n-n scattering more effective for 
vortex unpinning. 

Note that after successful unpinning, the vortices move radially outward. These unpinned vortices 
encounter a large number of vortices during their motion and should trigger 
an avalanche through the so-called \lq proximity knock-on\rq $~$ process \cite{wars12, wars13}. 
Thus, the vortex avalanche can be a viable process to produce large size glitches in this 
picture of local unpinning. We will explore the feasibility of an avalanche later in section 
\ref{section:avalanche}.
%%%%%%%%  FIGURE - 2 %%%%%%%%%
%%%%%%%%%%%%%%%%%%%%%%%
\section {Results : Numerical values of unpinned vortices and the glitch size}
\label{section:result}
%%%%%%%%%%%%%%%%%%%%%%%
\subsection{Unpinned vortices}
The volume of the pinned region depends on the availability of energy for deposition and a few other 
parameters such as $k_f$ (or, $n_f$), $E_f$ and $\Delta_f$. The value of the neutron superfluid Fermi 
momentum $k_f$ depends on the local mass density $\rho$, which increases 
with the depth of the inner crust \cite{epst96,past11}, and lies in the range $\sim (10^{11} 
- 10^{14})~\text{gm-cm}^{-3}$. Thus, the neutron density $n_f = k_f^3/3\pi^2$
also varies accordingly. We will use the results of \cite{past11}, 
where the Fermi momenta have been calculated at various values of the local mass density. For superfluid gap 
parameter $\Delta_f (k_f)$, we used the analysis of \cite{sinh15}, where the authors 
have numerically fitted their data to find the analytical expression of the gap parameter as a function of $k_f$ 
(Fig. \ref{fig:fig2}). Thus, knowing the variation of $N_v$ with $k_f$, supplemented with the information 
about the variation of $k_f$ with $\rho$, and the variation of $\rho$ with the depth of the inner crust, 
allow us to calculate the number of unpinned vortices at different location of the cylindrical shell.

\begin{figure}
\centering
\vspace{0.5cm}
\includegraphics[width=1.0\linewidth]{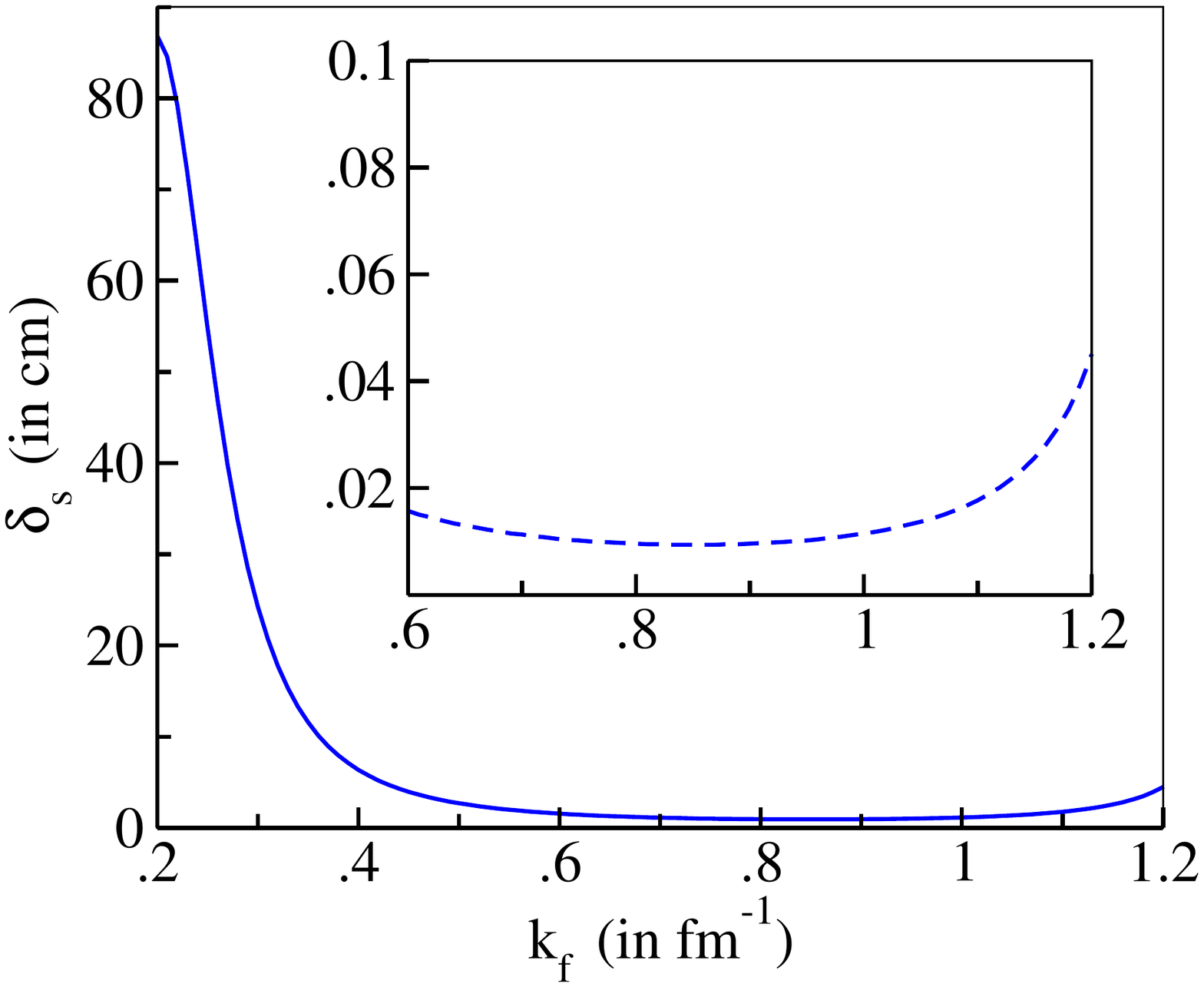}
\vspace{-0.9cm}
\caption{Thickness $\delta_s$ of the shell versus the Fermi momentum $k_f$.  
Inset shows the same plot with smaller scale for demonstration of the lower
part of $\delta_s$ more clearly.}
\label{fig:fig3}
\end{figure}

\begin{figure}
\centering
\vspace{0.5cm}
\includegraphics[width=1.0\linewidth]{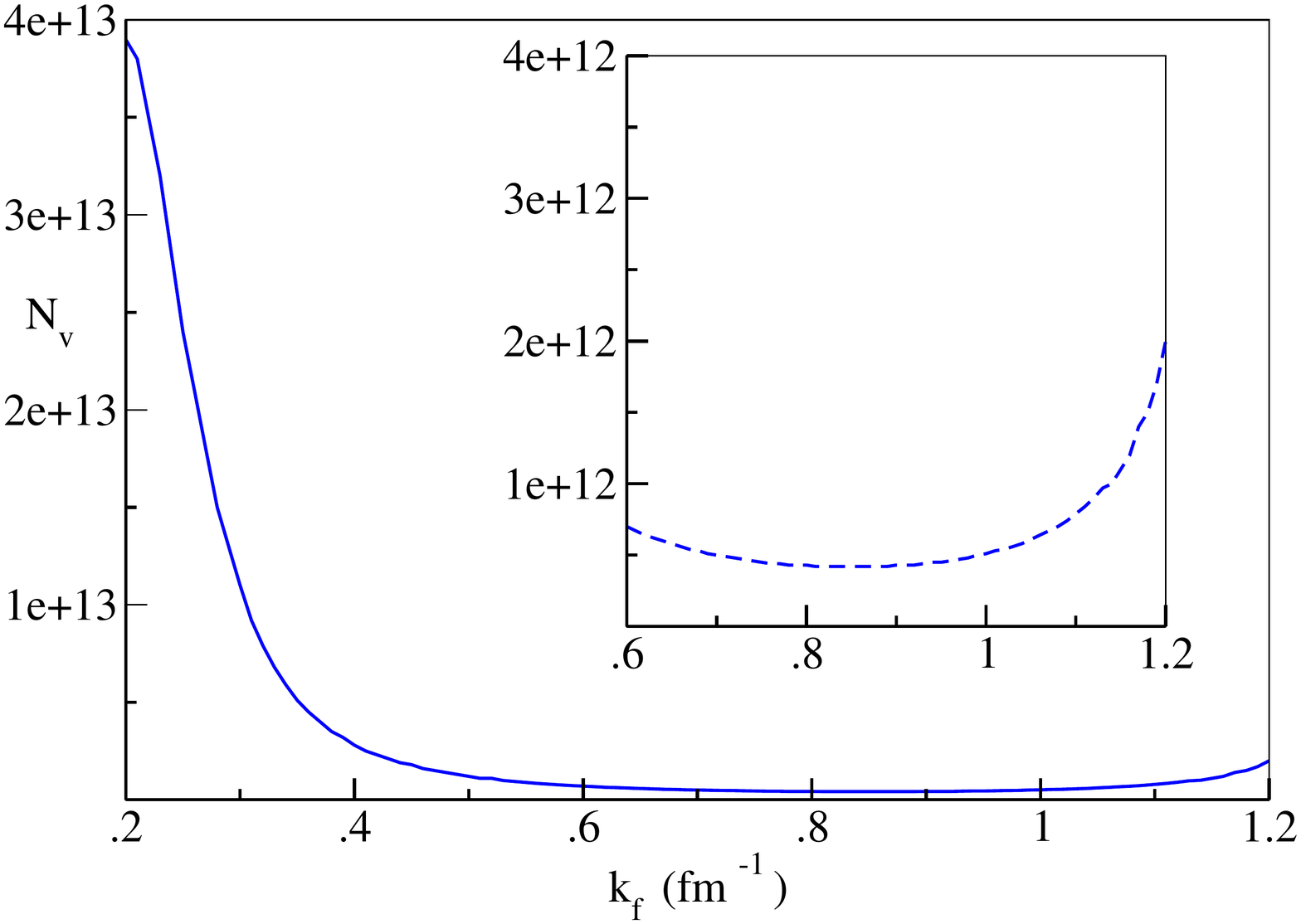}
\vspace{-0.5cm}
\caption{The number of unpinned vortices ($N_v$) caused by the neutron-vortex scattering 
in the cylindrical shell of thickness $\delta_s$ versus the Fermi momentum $k_f$.
For clarity, inset shows only the lower part of $N_v$.}
\label{fig:fig4}
\end{figure}

We have taken the values of $k_f$ from $0.2~\text{fm}^{-1}$ to $1.2~ \text{fm}^{-1}$ with the corresponding 
mass density $\rho$ in the range $\sim (10^{12} - 10^{14})~ \text{gm-cm}^{-3}$. Accordingly, the distance $R_s$ of 
the shell from the star's center is taken from $10.3$ km to $9.9$ km. The results are shown in 
Fig. \ref{fig:fig3} and Fig. \ref{fig:fig4}. The plots show that as $k_f$ increases 
from $0.2~\text{fm}^{-1}$ to $1.2~\text{fm}^{-1}$, the thickness $\delta_s$ of the shell, and hence 
the number of unpinned vortices $N_v$ decreases almost monotonically. The values of $\delta_s$ and $N_v$ 
lie in the range (87 - 1) cm  and $(10^{13}$ - $10^{12})$, respectively. Since $k_f$ maps
the depth of the crust, the variation of $\delta_s$ and $N_v$ are due to the location of the shell in our 
cylindrical geometry. For illustration, $k_f = 0.2$ corresponds to relatively outer layer of the inner 
crust ($R_s \simeq 10.3$ km) with mass density $\rho \simeq 10^{12}~ \text{gm-cm}^{-3}$. Therefore, the 
energy deposition around this region leads to the release of about $ 10^{13}$ vortices due to the scattering 
by the excited neutrons. Similarly, if the shell is located around the region with $k_f = 1.2$ 
($R_s \simeq 9.9$ km), the number of released vortices will be reduced to $ 10^{12}$. Although the number 
is reduced with the depth of the crust, this has an important implication in the context of vortex avalanche. 
We explore this possibility in section \ref{section:avalanche}.      
%%%%%%%%%%%%%%%%%%%%%%%
\subsection{Glitches} 
\begin{figure}
\centering
\vspace{0.5cm}
\includegraphics[width=1.0\linewidth]{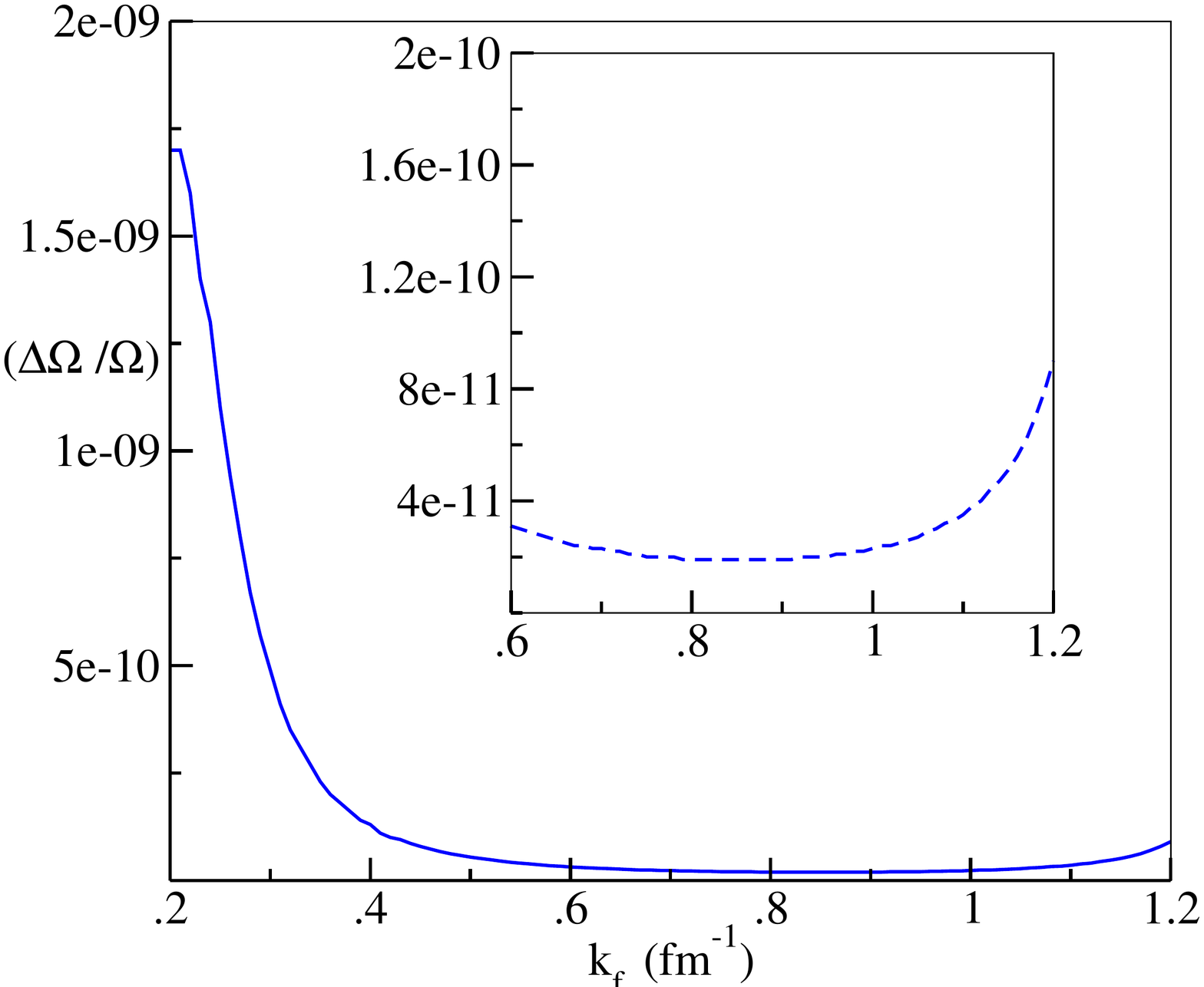}
%\vspace{-0.1cm}
\caption{The typical glitch size for a Vela like pulsars caused by the local unpinning from 
the cylindrical shell of thickness $\delta_s$ versus the Fermi momentum $k_f$.
For clarity, inset shows only the lower part of $\Delta \Omega/\Omega$.}
\label{fig:fig5}
\end{figure}

The glitch size can be determined using Eq. (\ref{eq:glitch}). The interglitch time $t_p$ 
is set by the frequency of occurrence of crustquake events and is proportional to the change of oblateness 
$\Delta \epsilon$. We choose $\Delta \epsilon = 10^{-8}$ to be consistent with the observed typical 
interglitch time $t_p \simeq 1$ year. It is now evident from the Eq. (\ref{eq:glitch}) that the glitch 
size of an individual pulsar of characteristic age $\tau$ depends on various neutron star parameters, 
such as $R$, $R_s$ and $\Delta R$, etc. The values of these parameters are taken by following the model 
of neutron star structure as mentioned in section \ref{section:basic}. The glitch size also depends on 
the ratio of the moment of inertia of the superfluid component in the inner crust to the rest of the star 
$I_f/I_c$. This ratio depends on a specific glitch model. For example, the author in Ref.\cite{rude76} 
suggested that a normal liquid layer can exist between the inner crust neutron superfluid and the core 
superfluid, resulting in a larger MI ratio. Without such a transition layer, the ratio can be of order 
$\sim 10^{-2}$ \cite{rude76}. There was also an attempt to fix this ratio 
through statistical analysis of glitches of Vela, Crab, and a few other pulsars (see Ref.\cite{link99}). 
Using the observed values of glitch parameters, the authors constrained the ratio as $I_{res}/I_c \ge 1.4\%$ 
for Vela. Here $I_{res}$ is the MI contribution by the angular momentum reservoir components. In our case, 
the whole inner crust superfluid is assumed to be an angular momentum reservoir, i.e., $I_{res} \equiv I_f$. 
It is then reasonable to take $I_f/I_c \simeq 10^{-2}$ for the estimate of glitch size for Vela. For 
consistency, note that by setting $\delta_s = \Delta R$ and $R \sim R_s$,  the above value of MI ratio 
results in maximum glitch size of order $10^{-6}$ for Vela pulsar ($\tau \simeq 10^{4}$ years). This is 
expected as the above glitch size corresponds to the release of all vortices. As noted in Ref.\cite{link99}, 
the analysis of Crab pulsar reveals a tiny MI ratio $I_f/I_c \simeq 10^{-5}$. It was suggested that losing 
angular momentum between glitches through vortex creep may be responsible for the lower value of the MI ratio. 
However, one should note that putting $I_f/I_c \simeq 10^{-5}, \delta_s =\Delta R$ (For the case when all the 
vortices are released.), $R_s \simeq R$ and $\tau \simeq 10^3$ years in Eq. (\ref{eq:glitch}), the smaller 
value of the MI ratio of Crab does not contradict the observed glitch size $\sim 10^{-8}$ of this pulsar. 
Thus, the above discussion suggests that the values of $I_f/I_c$, as mentioned above, are compatible with 
Vela and Crab pulsars. Now, the glitch size through local unpinning, where only a fraction of total 
vortices are released, depends on the value of $\delta_s$ and $\Delta R$.
In terms of these quantities, we get the glitch size of the Vela pulsar (with $\tau \simeq 10^4$ years) 
from Eq. (\ref{eq:glitch}) as
\begin{equation} \label{eq:micro}
\Big (\frac {\Delta \Omega}{\Omega}\Big) \simeq 10^{-6} \Big(\frac{\delta_s}{\Delta R}\Big).
\end{equation}
Thus, for $k_f$ in the range $(0.2 - 1.2)~ \text{fm}^{-1}$, the glitch size for the Vela pulsar 
lies in the range $(10^{-9} - 10^{-11})$ as shown in Table \ref{tab:table1} 
and in Fig. \ref{fig:fig5}. The above results follow from the local unpinning, without the 
effects of an avalanche. In the next section, we will discuss the implication of local 
unpinning on vortex avalanche. 

Before concluding this section, we should comment on our results' sensitivity 
to the neutron star model. In this study, we have taken a generic model for the internal 
structure of a neutron star and fixed the values of various neutron star parameters accordingly. 
The parameter-dependent sensitivity of our results can be understood from Eq. (\ref {eq:glitch}). 
As mentioned earlier, the MI ratio $I_f/I_c$ depends on the specific glitch model. Assuming the 
standard superfluid vortex picture (i.e., the inner crust superfluid component is the angular 
momentum reservoir responsible for the glitch event), the above ratio can be fixed following 
the statistical analysis of glitch events \cite{link99}. Once $I_f/I_c$ is specified, the glitch 
size should depend only on $\delta_s$ and $\Delta R$. Now $\delta_s(k_f)$ is a function of Fermi 
momentum $k_f$, which depends on the local baryon density in the inner crust region. The 
density-dependent Fermi momentum is now taken from Ref.\cite{past11}. As the methodology 
breaks down beyond $\rho \simeq 10^{14}~\text{gm~cm}^{-3}$, we restricted our study up 
to the baryon density $10^{14}~\text{gm~cm}^{-3}$ and the corresponding Fermi momentum  
$k_f = 1.2 ~\text{fm}^{-1}$. Note these calculations rely on the methodology used to study 
the properties of matter's superfluid at a high baryon density regime. Though the purpose of 
such a study is to understand the properties of the inner crust of a neutron star, the author 
\cite{past11} did not assume any specific model for neutron star structure. We adopt their 
data of density-dependent Fermi momentum for our purpose. For the inner crust density profile, we 
have followed the work of Ref.\cite{epst96}, where such a profile is provided for a neutron star 
of mass $1.4 M_\odot$ (see Ref.\cite{epst96} and the references therein for details). The value 
of $\delta_s$ is quite sensitive to the Fermi momentum and varies by about two orders of magnitudes, 
as $k_f$ changes from $0.2 ~\text{fm}^{-1}$ to $1.2 ~\text{fm}^{-1}$ (see Table \ref{tab:table1}). 
So finally, relying on the works of Refs. \cite{past11,epst96}, the model-dependent uncertainty 
in our results may arise only due to the value of the inner crust thickness $\Delta R$. For a given 
neutron star mass, the value of $\Delta R$ and the other neutron star parameters depend on the equation 
of state of the neutron star matter. For an of mass $\simeq 1.4 M_\odot$, the value of $\Delta R$ lies in the 
range (1-2) km \cite{epst96,mark96,link99, stei13}. Accordingly, we have taken $\Delta R = 1$ km while 
estimating the number of pinned vortices in the inner crust and the glitch size. Following the 
above discussion, we can safely say that our results are trustworthy for the pulsars of mass 
$\simeq 1.4 M_\odot$, i.e., the EOS-dependent sensitivity should not affect the order of magnitude 
estimate of glitch size.

%%%%%%%      TABLE 1  %%%%%%%%%
\begin{table}
\centering
\caption{Fermi momentum ($k_f$), distance of the cylindrical shell from the center ($R_s$), 
thickness of the shell ($\delta_s$), the number of unpinned vortices ($N_v$), and the order 
of magnitude of the glitch size ($\Delta \Omega/\Omega$) for the Vela pulsar through local unpinning by 
excited neutrons.}
\label{tab:table1}
\begin{tabular}{lcccc} % four columns, alignment for each
\hline
$k_f$ ($\text {fm}^{-1}$) & $R_s$ (in km) & $\delta_s$ (in meter) & $N_v$ & $(\frac{\Delta \Omega}{\Omega}) $   \\
\hline
0.2 & 10.3 & 0.87 & $3.9 \times 10^{13}$ & $~~~\sim 10^{-9}$ \\
\hline
0.8 & 10.2 &  0.01 & $ 4.3 \times 10^{11}$ & $~~~\sim 10^{-11}$ \\
\hline
1.2 & 9.9 & 0.05  & $2.0 \times 10^{12}$ & $~~~\sim 10^{-9}$ \\
\hline
\end{tabular}
\end{table}
%%%%%%%%  FIGURE -3 %%%%%%%%%
%%%%%%%%  FIGURE - 4 %%%%%%%%%
%%%%%%%%  FIGURE - 5 %%%%%%%%%
%%%%%%%%%%%%%%%%%%%%%%%
\section {The vortex avalanche through proximity knock-on}
\label{section:avalanche}
Instantaneous release of a large number of vortices ($\sim 10^{18}$) is necessary
for explaining large size glitches ($\Delta \Omega/\Omega \sim 10^{-6}$) of the pulsars. This 
seems to be feasible through the mechanism of vortex avalanche as suggested in the 
literature \cite{mela08, wars12, akba17}. For this to happen, a trigger mechanism is 
required to initiate the process like in any other natural avalanche event. In our model of 
local unpinning, the movement of the unpinned vortices may trigger an avalanche through the
proximity knock-on process \cite{wars13}. We will see below that the usefulness of such events 
from the context of explaining large size glitches depends on the 
position of the shell. The scope of knock-on is reduced if the shell is located 
very close to the outer part of the inner crust (Recall the statement in section \ref{section:intro} 
regarding the energy deposition in the inner crust.). For a qualitative understanding of the process, 
we follow the basic picture of proximity knock-on as described in Ref.\cite{wars13}. In the 
proximity picture, when an unpinned vortex comes closer to a pinned vortex from an intervortex 
distance $d_v$ to a distance $\eta d_v$ ($\eta <1$), it effectively reduces the pinning barrier. 
Here, the difference in angular frequency $\delta \Omega = (t_p/2\tau)\Omega$
(Eq. \ref{eq:domega1}) between the superfluid and the crust is modeled as a pinning barrier, which 
is expressed as 
\begin{equation} \label{eq:gamma}
\gamma = (1- \frac{\delta \Omega}{\delta \Omega_{cr}}) = 1 - (\frac{t_p}{2\tau}) 
(\frac{\Omega}{\delta \Omega_{cr}}). 
\end{equation}
The critical value of $\delta \Omega$ at which the magnus force balances the pinning force is denoted 
by $\delta \Omega_{cr}$, and is given by \cite{alpa84}
\begin{equation} \label{eq:domegacr}
\delta \Omega_{cr}= \frac{E_p}{\rho \kappa R b \xi}.
\end{equation}
Where, $\kappa = h/2m_n$ is the quantum vorticity with $m_n$ being the mass of a neutron.
The coherence length of the bulk superfluid is denoted by $\xi~(\simeq 10 ~\text{fm})$. The 
nucleus-nucleus distance is denoted by $b~ (\simeq 100~\text{fm})$, and $R$ ($\simeq 10$ km) 
is the (average) distance of the inner crust from the center of the star. For a given value of 
pinning energy $E_p$, the numerical value of $\delta \Omega_{cr}$ can be determined using 
Eq. (\ref{eq:domegacr}). The unpinning rate for a single vortex from a region characterized by 
the pinning energy $E_p$ is given by \cite{wars13}
\begin{equation} \label{eq:rate1}
\lambda_0 = \Gamma_0 e^{- \beta E_p \gamma}.
\end{equation}
The factor $\Gamma_0 = 10^{22}~\text {s}^{-1}$ is calculated by assuming the decay process to 
be of quantum origin (see Ref.\cite{wars13} and the references therein). The quantity $\beta (= 1/kT)$ 
characterizes the star's temperature. As an unpinned 
vortex gets closer to a pinned vortex, the pinning barrier is reduced by 
\begin{equation} \label{eq:barrier}
\Delta \gamma = (\frac{\kappa}{2\pi d_v R \delta \Omega_{cr}})(\frac{1-\eta}{\eta}), 
\end{equation}
and the unpinning rate is increased to 
\begin{equation} \label{eq:rate2}
\lambda = \lambda_0 e^{\beta E_p \Delta \gamma}.
\end{equation}
Note, Eq. (\ref{eq:rate1}) suggests that the unpinning rate from a specific region of the inner crust having a fixed 
value of $E_p$ is completely determined by the ratio of $\delta \Omega$ to $\delta\Omega_{cr}$. As the star slows down, 
$\delta \Omega$ gets close to $\delta \Omega_{cr}$. As a result, the $\gamma$ factor, which is interpreted as a pinning 
barrier, is decreased, causing the unpinning rate to increase with time. However, one should note that with a typical 
few years interglitch time $t_p$, $\gamma$ is approximately fixed. Also, as $\lambda$ varies with $E_p$, it's quite 
impossible that all the pinned vortices spread across the inner crust of the star get unpinned simultaneously. 
The vortex avalanche, therefore, seems to be a quite natural mechanism for unpinning of so many vortices. In our case, 
there are already a large number of unpinned vortices ($\sim 10^{14} - 10^{12}$) moving radially outward. 
These can act as triggers to 'knock-on' the pinned vortices in the inner crust of the star. 

Now,  Eq. (\ref{eq:rate2}) provides the vortex unpinning rate triggered by a single vortex. 
In our case, this will be modified due to the presence of a large number of triggers caused by 
local unpinning. For a qualitative understanding of 
the process, let us exploit the azimuthal symmetry of the vortex motion and focus only in one radial 
direction. We can estimate the effective number of triggers at the onset of the knock-on process across the 
shell of thickness $\delta_s$ with vortex 
density $n_v$ as $N^i_{tr} \simeq \delta_s ~\sqrt{n_v}$. As $\delta_s$ lies in the range $\sim (87 - 1)$ cm 
(depending on the location of the shell), $N^i_{tr}$ should lie in the range $\sim (10^3 - 10^2)\sqrt{\Omega}$. 
Here, $\Omega$ should be taken in units of $\text {s}^{-1}$. The value of $N^i_{tr}$, which is the effective 
numbers of triggers in one dimension sets the lower limit of triggers initiating the knock-on process. 
Eventually, these vortices may unpin the other vortices, which 
in turn should participate in the process. This increase the triggers $N_{tr}$ cumulatively. The above
picture should eventually lead to a few orders of magnitude enhancement of unpinning rate of a single vortex. 
The effects of these multiple triggers can be incorporated by modifying Eq. (\ref{eq:rate2}) as
\begin{equation} \label{eq:rate3}
\lambda_{tr} = N_{tr} \lambda = \Gamma_0 N_{tr} e^{-\beta E_p \gamma} e^{\beta E_p \Delta \gamma}.
\end{equation}
An average time $\tau_{tr}$ for a single unpinning event can be taken as $\tau_{tr} \simeq d_v/v_r$.
Where, $v_r \simeq R_s \delta \Omega \simeq (t_p/2\tau) \Omega R_s$ 
(see Eq. \ref{eq:domega1}) is the radial velocity of a vortex, and $d_v \sim 1/\sqrt{n_v}$ is 
the inter vortex distance. For the Vela pulsar with $t_p \sim 1$ year, $\tau_{tr}$ is about 
$10^{-7} $ s. For conservative estimate, let us set $N_{tr} = N^i_{tr}$. 
In this case, the unpinning probability for a single vortex can be written as
\begin{equation} \label{eq:prob}
\lambda_{tr} \tau_{tr} = \Big(\frac{2\tau }{t_p \Omega R_s \sqrt{n_v}} \Big)~N_{tr}~\Gamma_0~e^{-\beta E_p \gamma} 
e^{\beta E_p \Delta \gamma}.
\end{equation}
%%%%%%%      TABLE 2    %%%%%%%%%

\begin{table}
\centering
\caption{A Few values of the Fermi momentum $k_f$ at which the unpinning 
probability for a single vortex is enhanced due to multiple triggers.}
\label{tab:table2}
\begin{tabular}{lccc}
		\hline
		$k_f$ ($\text {fm}^{-1}$) & $\lambda \tau_{tr}$ & $\lambda_{tr} \tau_{tr}$   \\
		\hline
		$0.52$ & $\sim 10^{-3}$ &  $\sim$ 1.0  \\
                $1.14$ & $\sim 10^{-3}$ & $\sim$ 1.0  \\
                \hline
\end{tabular}
\end{table}
In the picture of proximity knock-on, unpinning occurs provided the condition $\lambda_{tr} \tau_{tr} \simeq 1$ 
is satisfied \cite{wars13}. In Table \ref{tab:table2}, we have listed a few values of the Fermi momentum $k_f$ 
around which the above condition is fulfilled in the presence of multiple triggers. 
For example, 
in the region around $k_f = 1.14~\text {fm}^{-1}$, $\lambda \tau_{tr} \simeq 10^{-3}$ and 
$\lambda_{tr} \tau_{tr} \simeq 1$.
Thus, there is about three order of magnitude enhancement in the unpinning probability for 
a single vortex in the presence of triggers. Similar effects are also observed around $k_f \simeq 0.52~\text 
{fm}^{-1}$. The multiple triggers could therefore significantly enhance the chances of unpinning from a few 
specific regions. However, it should be mentioned that there are also a few regions, where 
$\lambda_{tr} \tau_{tr}$ is still too low to meet the condition for unpinning.

We have described above a qualitative picture of the knock-on process. To properly 
understand the implications of avalanche, one has to resort to dynamical simulation. 
Also, one should note that the unpinning rate being depend on the exponential function of 
various parameters [Eq. (\ref {eq:rate2}) and/or Eq. (\ref{eq:rate3})], the results are quite sensitive 
to the values of those parameters. These factors should be taken into account in the proper analysis
of vortex avalanche. 

We can make a rough estimate for the expected glitch size, if we assume that the knock-on causes 
unpinning from a region of total thickness $\delta_{a}$. Here $\delta_{a}$ is the sum of 
the thickness of various regions where the unpinning condition is satisfied. This also includes 
the region, which were already affected by the local unpinning. So, the glitch 
size for the Vela pulsar will be modified as
\begin{equation} \label{eq:macro1}
\Big (\frac {\Delta \Omega}{\Omega}\Big) \simeq 10^{-6} \Big(\frac{\delta_{a}}{\Delta R}\Big) .
\end{equation}
The maximum value of $\delta_{a}$ is expected to occur, if the cylindrical shell is located 
around $R_s = 9.9$ km ($k_f = 1.2~\text {fm}^{-1}$). In this case, there are about $10^2 \sqrt{\Omega}$
triggers available (along a particular direction) for the process. These vortices while moving, unpin 
the vortices from the region whenever the quantity $\lambda_{tr} \tau_{tr}$ is approximately closed 
to unity. If all the vortices beyond $R_s = 9.9$ km get unpinned, the value of $\delta_{a}$ will be 
approximately equal to 400 meter (from $R_s = 9.9$ km to $10.3$ km). For this case, the glitch 
size will be of order $10^{-6}$ for a Vela-like pulsars (with $\tau \sim 10^4$ years). However, analyzing 
the unpinning probability at various regions (a few regions have too low probability), only a fraction $f_{a} 
= (\delta_{a}/400~\text{m})$ of the region beyond $R_s = 9.9$ km seems to be affected by the trigger 
mechanism. For proper estimate of $f_{a}$, a proper numerical algorithm is required to implement 
the above process (see Ref.\cite{wars13} for one such algorithm). It will be then interesting to 
see if the local unpinned vortices trigger an avalanche to produce a large size glitches without affecting 
the typical one year interglitch time. Note, in a crustquake model, the larger size glitches need a larger 
interglitch time \cite{baym71} contrary to the observations.  
%%%%%%%%%%%%%%%%%%%%%%%
\section {Comment and Conclusion}
\label{section:conclusion}
We have observed that the crustquake followed by vortex unpinning from a cylindrical shell may produce 
glitches of size $10^{-9} - 10^{-11}$ for relatively old pulsars with characteristic age $\tau \sim 10^4$ 
years. There is also scope for the generation of larger size glitches through avalanche triggered by the 
unpinned vortices as observed by the Vela pulsar. Although not regular, 
the Crab pulsar also exhibited a large size glitch ($\Delta \Omega/\Omega \sim 0.52 \times 10^{-6}$) 
as reported in 2017 \cite{shaw18}. It is impossible \cite{shaw18} to explain such a large glitch 
from a younger pulsar through the standard crustquake model. A partial avalanche (i.e., affecting 
only a few fractions of total vortices in the inner crust) triggered by the locally unpinned vortices 
can account for such occasional glitch activity of a young pulsar.

Besides the occurrence through vortex unpinning, the crustquake itself also produces glitches 
due to a sudden change of the pulsar's shape. The non-observation of successive glitches within a short 
span can be explained if the time interval between these two events turns out to be small compared to the 
current observational limit of individual glitches \cite{asht19}.
Now we estimate the time difference between these two glitch events. For that, we assume 
the energy deposition to the inner crust is almost instantaneous (See Ref.\cite{epst96} for the time 
evolution of the temperature profile in the inner crust after energy absorption.) and ignore the time 
scale associated with the vortex unpinning (n-n scattering time scale $\tau_{nn} \simeq 10^{-5}~ 
\text{s}$ only). Thus the time interval between two glitch events is determined by the time $t_v$ 
taken by the vortex to reach the outer crust. This can be obtained from the radial velocity $v_r$ 
of the unpinned vortices. For $v_r \simeq R (\delta \Omega) = (t_p/2\tau)~ \Omega R \simeq 10^4~ 
\text{cm-s}^{-1}$, the time $t_v \simeq v_r/\Delta R$ turns out to be 
$\sim 0.1 ~\text{s}$ for the Vela pulsar. The above time difference is associated with the 
glitches occurring through crustquake followed by the glitch produced through local unpinning. 
For the case of avalanche, a single unpinning event takes an average time $\tau_{tr} 
\simeq d_v/v_r \sim 10^{-7}~\text{s}$ and the whole process is expected to be completed within 
$\sim 10^{-6}~\text{s}$ (see \cite{prad1} and the reference therein). Thus we see that the time 
interval between the glitches produced by the crustquake and the vortex unpinning is of order 
tenth of a second, i.e., they seem to overlap. The short time interval also 
justifies our earlier assumption (in section \ref{section:basic}) that the waiting time
between two successive crustquake events almost overlaps with the time duration  $t = 0$ 
to $ t = t_p$, i.e., the time during which vortices remain pinned to the sites.

Thus, unifying crustquakes with the superfluid-vortex model can consistently produce regular glitches 
with a typical frequency of once every few years (set by the time interval of successive crustquake events). 
In case of an avalanche, unifying the models can produce large size Vela or Vela-like glitches without 
affecting the waiting time of crustquakes. Recall that the crustquake model alone is not compatible with 
such glitches. For Crab-like younger pulsars,  partial avalanche may be responsible for the occasional 
larger size glitch activity \cite{shaw18}. 

From the observational perspective, as we mentioned earlier, with the current resolution \cite{asht19}, 
it is impossible to resolve the subsecond time interval between the glitches produced by the crustquake 
followed by a glitch through unpinning. For the case of vortex avalanche, though, it's the larger size 
glitch that is expected to dominate the glitch feature. Hence the source of the larger glitch (i.e., vortex 
unpinning) will be easily identifiable. However, if the avalanche mechanism is ineffective (for example, 
if the energy is deposited at the outer part of the inner crust), it will be an observational challenge 
to resolve two successive, almost identical size glitches . This is a common feature of all the proposed 
models, where both the crustquake and vortex model are involved in the generation of glitches (see, for 
example, the Refs. \cite{rude91c, epst96, akba17}). For such cases, the improved glitch observation time 
of individual pulsars can help identify the precise source of the glitches.

To conclude, we studied the vortex unpinning mechanism by scattering excited neutrons with the 
vortex core neutrons in the inner crust of the pulsars. The strain energy released by the crustquake 
is assumed to be absorbed in some part of the inner crust around the equatorial plane of the star 
and causes excitation of the free superfluid neutrons surrounding the vortices. The excited neutrons 
unpin a large number of vortices from that region and result in pulsar glitches. We considered a 
cylindrical shell around the equatorial plane and studied the effects of the thermally excited 
neutrons in unpinning the superfluid vortices from the affected site. As the affected pinned 
region's precise location is unknown, we took the shell at various depths (equivalently, at 
different local mass density regions) of the inner crust. This was achieved by varying the 
Fermi momentum $k_f$ of the superfluid neutrons from $0.2~\text{fm}^{-1}$ to $1.2~ \text{fm}^{-1}$.
We then determined the shell thickness $\delta_s$, which has been observed to lie in the range 
$(85 - 1)$ cm. 
The corresponding values of unpinned vortices caused by the neutron-vortex scattering were 
found to lie in the range $(10^{13} - 10^{11})$, which are equivalent to glitch size in the 
range $(10^{-9} - 10^{-11})$ for the Vela pulsars.

We suggested the possibility of a vortex avalanche triggered by the vortices, which were already 
unpinned from the cylindrical shell. In the picture of proximity knock-on, the presence of multiple 
triggers can enhance the unpinning probability and hence the glitch size. A rough estimate of glitch 
sizes has been presented. The result is quite encouraging from the perspective of producing a large 
size glitch without affecting the waiting time of successive crustquakes. The various time scales 
associated with our model are compatible with the observed features of glitches.\\
%%%%%%%%%%%%%%%%%%%%%%%

\section{Acknowledgements}
We thank the anonymous reviewer for critical comments and constructive suggestions on the previous 
version of this manuscript. 
%%%%%%%%%%%%%%%%%%%%%%%
%\bibliographystyle{apsrev4-2}
%\bibliography{glitch}
%%%%%%%%%%%%%%%%%%%%%%%
%apsrev4-2.bst 2019-01-14 (MD) hand-edited version of apsrev4-1.bst
%Control: key (0)
%Control: author (8) initials jnrlst
%Control: editor formatted (1) identically to author
%Control: production of article title (0) allowed
%Control: page (0) single
%Control: year (1) truncated
%Control: production of eprint (0) enabled
%
\end{document}